\documentclass[journal,letterpaper]{IEEEtran}

\usepackage{graphicx,cite,epsfig,amssymb,amsmath,amsfonts,multicol,subfigure,mathtools,bm,mathrsfs,setspace}
\usepackage{multirow}
\usepackage{xcolor}

\begin{document}
\title{Application of Opportunistic Bit to Multilevel Codes}
\author{
	Bingli Jiao, Mingxi Yin and Yuli Yang

\thanks{Bingli Jiao and M. Yin are with the Department of Electronics, Peking University, Beijing 100871, China (e-mail: jiaobl@pku.edu.cn, yinmx@pku.edu.cn).}
\thanks{Y. Yang is with the School of Engineering, University of Lincoln, Lincoln, U.K. (e-mail: yyang@lincoln.ac.uk).}
}

\maketitle
       
\begin{abstract}
In this paper, we propose a new signal organization method to work in the structure of the multi level coding (MLC).  The transmit bits are divided into opportunistic bit (OB) and conventional bit (CB), which are mapped to the lower level- and higher level signal in parallel to the MLC, respectively. Because the OB's mapping does not require signal power explicitly, the energy of the CB modulated symbol can be doubled.  As the result, the overall mutual information of the proposed method is found higher than that of the conventional BPSK in one dimensional case.  Moreover, the extension of the method to the two-complex-dimension shows the better performance over the QPSK.   The numerical results confirm this approach.

\end{abstract}


\IEEEpeerreviewmaketitle

\section{Introduction}
The idea of the multi level code (MLC) is to partition a signal set consisting of the high order modulation into the multiple levels, each of which contains low order modulation with the binary coded signals[1][2].  The advantage lies in the fact that the partitioned signal sets can use different channel codes to satisfy their individual error protections.  However, the overall information of the conventional MLC methods is found the same as that of the original signal set without any partitioning.    

In contrast to the signal partition, this paper proposes a new method that define two kinds of channel code-bits and organizes them into two-level-structure of the MLC. One of the two kinds uses the mapping method based on concept of the opportunistic bit (OB) and the other uses the conventional modulation.  The method of the OB was originally introduced for increasing the spectral efficiency of TCP/IP transmission, where the OB's mapping does not use explicitly either time resource and signal power, but uses the induces of the pre-designed time slots (TSs)~\cite{Jiao_TVT}~\cite{Yin2019}.  Actually, the saving of time resource can increase the bandwidth efficiency and that of signal power the energy efficiency.  The gains of the both are the same as described by $\eta= (K_c+K_o)/K_c$, where $\eta$, $K_c$ and $K_o$ are the gain factor and the numbers of OBs and CBs in a data unit, respectively.  The weakness is, however, at its dependency on the use of some information bits embedded in each header of the IP packet.    

In order to leverage OB concept to save the signal power in ML structure,  we categorize the transmit bits into OB and CB to form low-level- and high level signal, where the former is mapped onto a switching that selects the latter or a vacant symbol, i.e., a symbol without any power, into Euclidean space.  To be specific, when bit value of the OB equals 1, a BPSK mapped by CB is switched to a transmission, while bit value of OB 0,  a vacant symbol to the transmission.  The construction of the proposed two-level signals is completed with OB transmitted at every symbol duration and CB randomly depending on the OB's value. We refer the proposed method to as the OB-assisted multi-level coding (OB-MLC) Scheme, which will be developed in one- and extended to two dimensional case for the comparison with BPSK and QPSK respectively. 

The remainder of this paper is organized as follows.  Section II sets up the system model, formulates the mathematical problem of the mutual informations and makes the comparisons with BPSK and QPSK in one- and two dimensional case, respectively.  The conclusion is given in section III.     

Throughout the paper, the following notations are used:  Capital Letter represents the vector in Hamming space, e.g., $ V = [v_1,v_2,\cdots,,v_n,\cdots,v_N] $, where $V$ is the bit vector and $v_n \in  \{1,0\}$ is the $n^{th}$ component,  and bold face letter the vector in Euclidean space, e.g., ${\bf{x}}=[x_1,x_2,\cdots,x_n,\cdots,x_N]$, where ${\bf{x}}$ is the signal vector and $x_n$ is the component, respectively.  

\section{Signal Scheme and Numerical Result}
Assume that we are working at the two dimensional complex additive white Gaussian noise (AWGN) channel
\begin{equation}\label{1-1}
{\bf{z}} = {\bf{x}} + {\bf{n}}
\end{equation}  
with $j=\sqrt{-1}$, where ${\bf{z}} = z^i + jz^q$ channel output containing the in-phase term $z^i$ and quadrature term $z^q$, ${\bf{x}}=x^i+jx^q$ are channel input with the in-phase signal $x^i$ and quadrature signal $x^q$,  and ${\bf{n}}=n^i+n^q$ are the Gaussian noise with the in-phase- and quadrature component, with the noise powers $n^i \sim \mathcal{N}(0,{{\sigma^i}^2})$ and $n^q \sim \mathcal{N}(0,{{\sigma^q}^2} )$, respectively.   

The OB-MLC method is explained first in one dimensional case, i.e., the co-phase team and, then, extended to the two dimensional case by adding the quadrature term into account.  

\subsection{One Dimensional Case}    
Let us work on the in-phase component of \eqref{1-1} and consider two independent sets of binary channel code-bits, marked by $V^c = \{v^c_1,v^c_2,\cdots,v^c_{m_c},\cdots,v^c_{M_c}\}$ and $ V^o= \{v^o_1,v^o_2,\cdots, v^o_{m_o},...,v^o_{M_o}\}$,
with $v^c, v^o \in 0,1 $ and $M_c=M_o/2$, to be mapped onto  two-level signals, where $V^c$ is defined as the CB and $V^o$ the OB ,  respectively.   

\begin{figure}
	\centering
	\subfigure[Equivalent channel for the high level if $v^o=1$]{
		\includegraphics[width=0.42\textwidth]{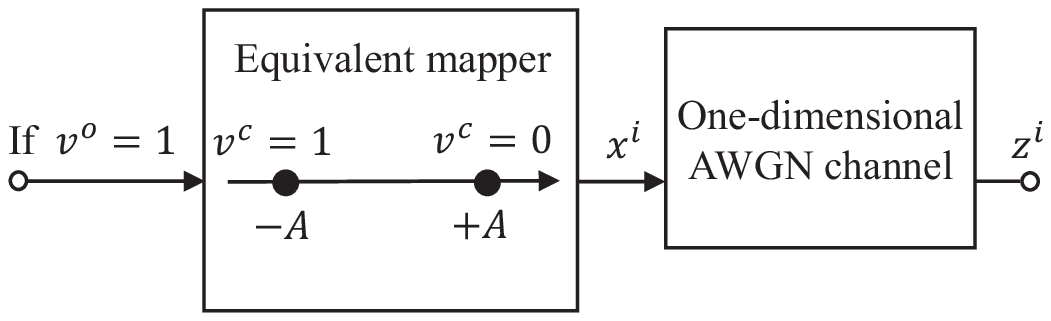}
		\label{fig1a}}
	\subfigure[Equivalent channel for the high level if $v^o=0$]{
		\includegraphics[width=0.42\textwidth]{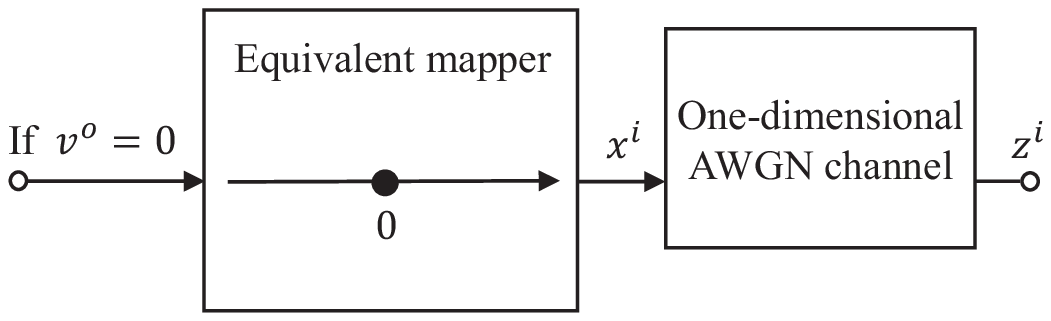}
		\label{fig1b}}
	\subfigure[Equivalent channel for the low level]{
		\includegraphics[width=0.42\textwidth]{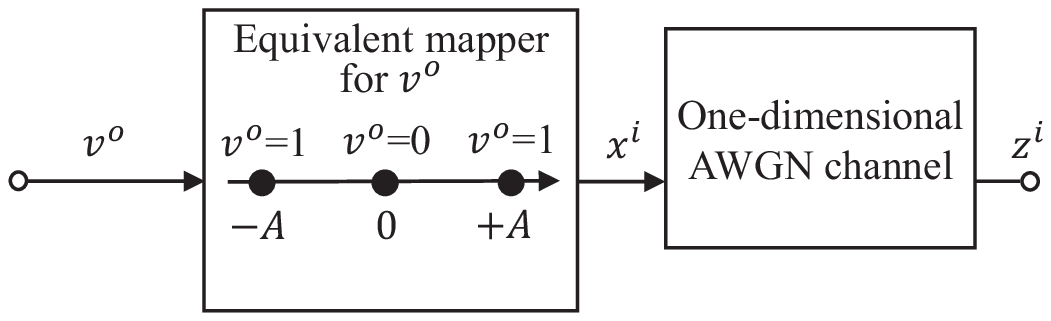}
		\label{fig1c}}
	\caption{One-dimensional mapping between multi-level codes and the constellation.}
	\label{fig_1Dsys}
\end{figure}

First, we modulated each bit of the CB onto a BPSK symbol  sequentially as the high-level signal  
\begin{equation}\label{1-2}
\zeta_{m_c}= A(1-2v^c_{m_c}) \ \ \ \  \text{for} \ \ m_c=1,2,...M_o/2  
\end{equation}
with $\zeta_{m_c} = +A$ and $-A$ representing $v^c_{m_c} =0$ or $1$ in Euclidean space respectively, where  $A$ is the amplitude of the BPSK and $\zeta_{m_c}$ is defined as the ${m_c}^{th}$ CB-symbol.  

It is noted that ${\bf{\zeta}}=[\zeta_1,\zeta_2,...\zeta_{m_c},... \zeta_{M_c}] $ is the high-level signal set according to the MLC of two-level. 

Secondly,  the bit value of an OB is defined, sequentially, by switching a component in $\zeta$ or add vacant symbol, i.e. a symbol without any energy, to a channel realization as follows. If the OB's bit value equals $1$, the CB-symbol is transmitted. while, if it equals $0$, a vacant symbol is transmitted as shown in Fig. \ref{fig_1Dsys}, where the sub-figure (a) shows the case of $v^o=1$, in which a CB-symbol is transmitted with -A or A, and the sub-figure (b) the case of $v^o=0$, in which the vacant symbol is transmitted.  The low-level signal is shown in Fig. \ref{fig_1Dsys}(c) and the high-level is simply a BPSK.  In summarisation, the OBML transmits 1.5 bits per symbol duration.  

To formulate the signal sets in parallel to ML, the transmit signals can be expressed as
\begin{equation}\label{1-3}
x^i_{m_o} =
\begin{cases}
\zeta_{m_c} &\mathrm{if} \ v^o_{m_o}=1;\\  
0,  \qquad &\mathrm{if} \ v^o_{m_o}=0.
\end{cases} 
\end{equation}
with $ {m_c} = \sum v^o_{m_o}$ for $m_o=1,2,\cdots,....M_o$, where  $x^i_{m_o}$ is the ${m_o}^{th}$ transmit signal.  

We note that ${\bf{x}}^i = [x^i_1,x^i_2,.....,x^i_{m_o}, ....x^i_{M_o}]$ is the low-level signal set, of which the de-mapping of the OB value $v^o_{m_o}$ is to estimate if $ x^i_{m_o} $ belongs to $\zeta_{m_c}$ or $0$. While, the demodulation of high-level signals are done on the CB symbols by estimating if $\zeta_{m_c}=A$ or $\zeta_{m_c}=-A$. 

In calculating the overall mutual information of the OBML, we remove the sequential index $m_o$ and $m_c$ for formulating the problem in general form   
\begin{equation}\label{1-4}
z^i = x^i + n^i
\end{equation}  
where $z^i$ is the channel output of in-phase part of \eqref{1-1},  $x^i$ can be either a BPSK symbol or a vacant symbol with equal probability and $n^i$ is the Gaussian noise. 

The signal detection starts from the low-level signal, i.e. de-mapping the OB.  The Log likelihood Ratio (LLR) can be calculated by 
\begin{equation}\label{1-5}
\begin{aligned}
{\rm{LLR}}_o^i &= \ln \frac{{\sum\limits_{v_m^o = 0} {\exp \left( { - \frac{{{{\left\| {{z^i} - {x^i}} \right\|}^2}}}{{2\sigma _i^2}}} \right)} }}{{\sum\limits_{v_m^o = 1} {\exp \left( { - \frac{{{{\left\| {{z^i} - {x^i}} \right\|}^2}}}{{2\sigma _i^2}}} \right)} }} \\
&= \ln \frac{{2\exp \left( { - \frac{{{{\left\| {{z^i} - 0} \right\|}^2}}}{{2\sigma _i^2}}} \right)}}{{\exp \left( { - \frac{{{{\left\| {{z^i} - A} \right\|}^2}}}{{2\sigma _i^2}}} \right) + \exp \left( { - \frac{{{{\left\| {{z^i} + A} \right\|}^2}}}{{2\sigma _i^2}}} \right)}}
\end{aligned}
\end{equation}
where ${\rm{LLR}}_o^i$ is a soft decision of $v^o$.   

Examining \eqref{1-3}, we can find that the CB-symbol is transmitted in parallel with OB only when $v^o=1$.  Thus, before demodulating the CB, receiver need to decode $V^o$ and use the result $\hat{V}^o$ to find CB-symbol, where $\hat{V}^o$ is the decode result of the channel code.  

Then, the demodulation of CB is done sequentially over each BPSK symbol corresponding to $v^o=1$.  The LLR can be calculated by  
\begin{equation}\label{1-6}
{\rm{LLR}}_c^i = \ln \frac{{\exp \left( { - \frac{{{{\left\| {{z^i} - A} \right\|}^2}}}{{2\sigma _i^2}}} \right)}}{{\exp \left( { - \frac{{{{\left\| {{z^i} + A} \right\|}^2}}}{{2\sigma _i^2}}} \right)  }}
\end{equation}
where ${\rm{LLR}}_c^i$ is a soft decision of $v^c$. 

In comparison with the conventional BPSK, the OBML method can take the following two advantages.  First, the CB-symbol can use the doubled symbol energy as
\begin{equation}\label{1-7}
E^c = 2\bar{E^i}
\end{equation}
where $E^c$ is energy of one CB-symbol and $E^i$ is the average symbol energy of the signals, because a half number of the transmit symbols are the vacant symbol without energy.  Consequently, the half  number of the CB-symbols uses the total symbol-energy.    

Secondly, in comparison with conventional BPSK, the OBML method can gain a bandwidth efficiency in terms of bits per symbol by   
\begin{equation}\label{1-8}
\alpha = \left(1 + \frac{1}{2}\right)/1=1.5 
\end{equation}  
where $\alpha$ is the gain factor.  

The first term in bracket of \eqref{1-8} represents the bit number of OB and the second that of CB per symbol in average.   

Applying the Chain rule of mutual information [40, p22], the overall mutual information can be calculated by 
 
\begin{equation}\label{1-9}
I(X^i;Z^i) =  I(v^o;Z^i) + \frac{1}{2} I(v^c;Z^i|v^o)
\end{equation}  
where $I(X^i;Z^i)$ is the overall mutual information of the in-phase signal,  $I(v^o;Z^i)$ and $I(v^c;Z^i|v^o)$ are the mutual information owning to transmission of OB and the conditional mutual information  owing to that of CB, respectively.  

According to the definition in theorem \cite{Shannon1948}, 
the mutual information owing to the OB transmission can be calculated by 
\begin{equation}\label{1-10}
\begin{aligned}
&I(v^o;Z^i) = {\log _2}2 \\
&\qquad \qquad \quad-  \frac{1}{2}\sum\limits_{v^o=0,1} {\int {p\left( {{Z^i}|{v^o}} \right)\log \frac{{\sum\limits_{v_l^o=0,1} {p\left( {{Z^i}|v_l^o} \right)} }}{{p\left( {{Z^i}|{v^o}} \right)}}} } dy\\
&= I_o(\gamma^i)={\log _2}2 \\
&\quad - \frac{1}{2}{{\mathbb E}_W}\left\{ {\log \left( {1 + \frac{{f\left( {W,\sqrt {\gamma^i} } \right) + f\left( {W, - \sqrt {\gamma^i} } \right)}}{{2f\left( {W,0} \right)}}} \right)} \right\} \\
&\quad - \frac{1}{4}{{\mathbb E}_W}\left\{ {\log \left( {1 + \frac{{2f\left( {W, - \sqrt {\gamma^i} } \right)}}{{f\left( {W,0} \right) + f\left( {W, - 2\sqrt {\gamma^i} } \right)}}} \right)} \right\} \\
&\quad - \frac{1}{4}{{\mathbb E}_W}\left\{ {\log \left( {1 + \frac{{2f\left( {W,\sqrt {\gamma^i} } \right)}}{{f\left( {W,0} \right) + f\left( {W,2\sqrt {\gamma^i} } \right)}}} \right)} \right\}
\end{aligned}
\end{equation} 
where $\gamma^i= \bar{E}^i / {{\sigma^i}^2}$ denotes symbol energy in average to that of noise of the in-phase part, $E_c^i=2\bar{E}^i$ is the symbol energy of a CB and $\bar{E}^i=1/4\times (-A)^2+1/2\times 0^2+1/4\times A^2=A^2/2$ is the averaged symbol energy by assumption that $v^o$ is transmitted with equal probability at $v^o=0$ and $v^o=1$, and ${\mathbb E}[\cdot]$ denotes the expectation operator.  The function $f(W,a)$ in \eqref{1-9} is given by
\begin{equation}\label{1-11}
f(W,a) = e^{- {\left( W - a \right)}^2} 
\end{equation} 
and $W$ denotes a Gaussian variable with zero-mean and variance $1/2$, i.e., $W \sim \mathcal{N}(0,1/2)$.

Noting that $v^o$ is transmitted with equal probability at $v^o=0$ and $v^o=1$, we can calculate the mutual information owing to CB by  
\begin{equation}\label{1-12}
\frac{1}{2} I(v^c;Z^i|v^o)  = \frac{1}{2}I_B\left({E}^c/{{\sigma^i}^2}\right)  = \frac{1}{2}I_B\left(2\bar{E^i}/{{\sigma^i}^2}\right) 
\end{equation}  
where $I_B(\cdot) $ is the mutual information of the conventional BPSK.   

Taking \eqref{1-10} and \eqref{1-12} into \eqref{1-9} completes the derivations of the overall mutual information of  the OBML in one dimensional case. 

The overall mutual information can be calculated by \eqref{1-9} and the numerical results are shown in Fig. 2, where the performance of the OBML is found better than that of conventional BPSK, i.e., $I_B(\bar{E}^i/{\sigma^i}^2)$.  In addition, the results of \eqref{1-10} and \eqref{1-12} are presented for showing the contributions of the low-level- and high-level signal.  

To show the gain of the OB-MLC more clearly, the difference of the mutual informations between the one-dimensional OB-MLC and the conventional BPSK are plotted in Fig. \ref{fig_rate_1D_diff}, where one can find the gain is always larger than zero and increases rapidly with the SNR $\bar{E^i}/{{\sigma^i}^2}$ to its saturation value 0.5bit/s/Hz. 

\begin{figure}
	\centering
	\includegraphics[width=0.5\textwidth]{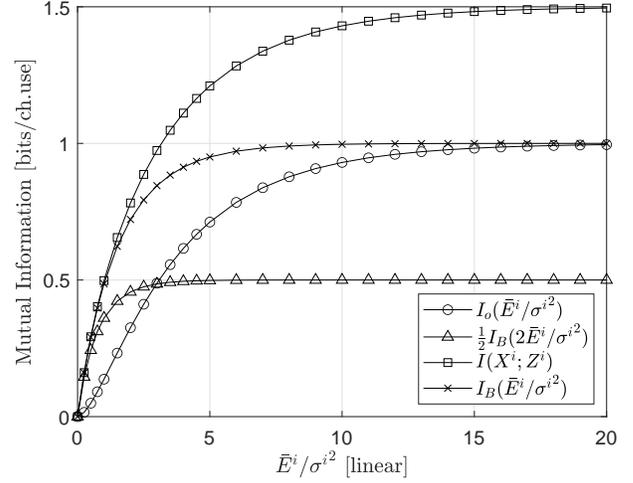}
	\caption{Mutual information for the one dimensional case.}
	\label{fig_rate_1D}
\end{figure}

\begin{figure}
	\centering
	\includegraphics[width=0.5\textwidth]{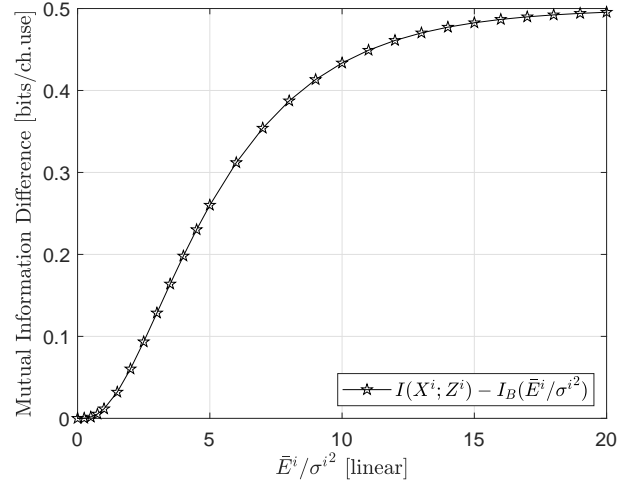}
	\caption{Difference between the mutual information of the one-dimensional OB-MLC and BPSK}
	\label{fig_rate_1D_diff}
\end{figure}


\subsection{Two Dimensional Case}
This subsection extends the OB-MLC to the two dimensional case by adding the quadrature component of \eqref{1-1} to complete the transmission over the complex channel model. 

In assumption of the same transmit signal power as that of the in-phase signal, the overall mutual information of the quadrature part is calculated.  Then, the summation of the mutual informations owing co-phase- and quadrature signal is defined as the total mutual information for the comparisons with that of the conventional QPSK. 

Because that the quadrature signals are orthogonal to the in-phase ones,  we can employ the scheme of the latter to the former in bits' transmissions explained as follows.  

Assume that the quadrature signals transmit their CB and OB marked by $ P^c= \{p^c_1,p^o_2,\cdots, p^c_{m_o},...,p^c_{M_o}\}$ and $Q^o = \{q^o_1,q^o_2,\cdots,q^o_{m_c},\cdots,q^o_{M_c}\}$ with an assumptions of $p^c_{m_c}, q^o_{m_o} \in 1,0 $  and $M_c=M_o/2$, respectively.  

The CB is also modulated by BPSK sequentially as   
\begin{equation}\label{2-1}
	\xi _{m_c}= A(1-2p^c_{m_c}) \ \ \ \    
\end{equation}
with $\xi _{m_c} = -A$ or $+A$ for $u^c_{m_c} =0$ or $1$ accordingly, where $A$ is the amplitude of BPSK and $\xi_{m_c}$ is defined as the ${m_c}^{th}$ CB-symbol in the quadrature part.  

In the parallel procedures of the in-phase signals, the transmit signal of the quadrature part can be expressed by  
\begin{equation}\label{2-2}
x^q_{m_o} =
\begin{cases}
\xi_{m_c},  \qquad &\mathrm{if} \ q^o_{m_o}=1;\\  
0 &\mathrm{if} \ q^o_{m_o}=0.
\end{cases} 
\end{equation}
with $ m_c = \sum u^o_{m_o}$ for $m_o=1,2,\cdots,M_o$.

The general form of quadrature-signal's transmission can be written as   
\begin{equation}\label{2-3}
z^q = x^q + n^q
\end{equation}  
where $z^q$ is the channel output of the quadrature part, $x^q$ is the quadrature signal which can be either a BPSK symbol or a vacant symbol with equal of each other and $n^q$ is the Gaussian noise.   

The overall mutual information owning to the quadrature signal is exactly same as that of the in-phase signal because \eqref{2-3} is essentially same as \eqref{1-4} with ${\sigma^i}^2={\sigma^q}^2$.  

Then, the total mutual information of the two dimensional OB-MLC can be calculated by 
\begin{equation}\label{2-4}
I({\bf{X};\bf{Z}}) =  2I_o\left(\bar{E}^i/{\sigma^i}^2\right) + I_{B}\left(2\bar{E}^i/{\sigma^i}^2\right)
\end{equation}  
where $I({\bf{X};\bf{Z}})$ is the total mutual information.  

Equation \eqref{2-4} can be written in the two dimensional case  
\begin{equation}\label{2-5}
I({\bf{X};\bf{Z}}) =  2I_o\left(\bar{E}/\sigma^2\right) + I_B\left(2\bar{E}/\sigma^2\right)
\end{equation} 
with $\sigma^2 =  {\sigma^i}^2 + {\sigma^q}^2$ and $\bar{E} = 2\bar{E}^i$, where $\bar{E}$ is the average symbol energy of two dimensional signals.

\begin{figure}
	\centering
	\includegraphics[width=0.5\textwidth]{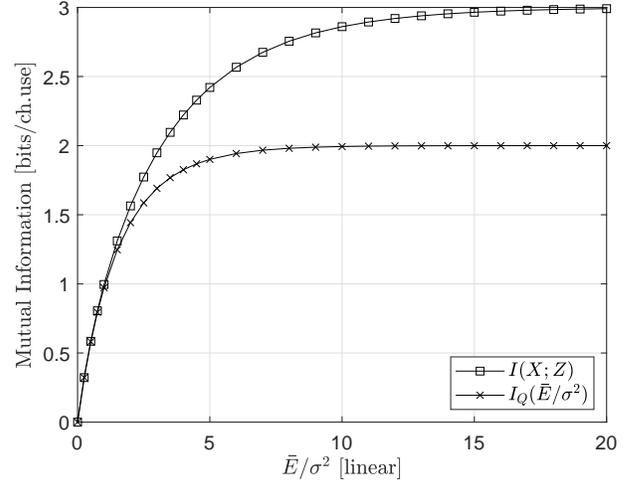}
	\caption{Mutual information for the two dimensional case.}
	\label{fig_rate_2D}
\end{figure}

The numerical results of \eqref{2-5} are shown in Fig. \ref{fig_rate_2D}, where one can find that the mutual information of the OB-MLC is larger than that that of the conventional QPSK.  

\begin{figure}
	\centering
	\includegraphics[width=0.5\textwidth]{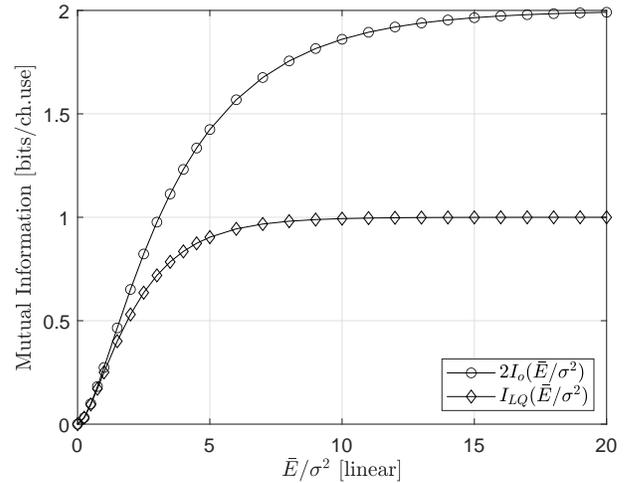}
	\caption{Mutual information for the low level comparison.}
	\label{fig_rate_2DL}
\end{figure}  

To explain the exceeding mutual information of the OB-MLC over the conventional QPSK above, we adopt the MLC of QPSK as a reference for the following reasons; (1) its overall mutual information is the same as that of the conventional QPSK and (2) its structure allows  the comparisons on the low-level and high-level signal in parallel to the MLC.  

According to ~\cite{MultilevelCodes}, the mutual information of the low-level signal of the conventional MLC for QPSK can be calculated by 
\begin{equation}\label{eq_QMI_L}
	\begin{array}{l}
		I_{LQ}(\gamma)= 1 \\
		-\frac{1}{4}{\mathbb{E}_{W_2}}\left[ {\log \left( {1 + \frac{{f\left( {{W_2}, - \sqrt \gamma   + j\sqrt \gamma  } \right) + f\left( {{W_2}, - \sqrt \gamma   - j\sqrt \gamma  } \right)}}{{f\left( {{W_2},0} \right) + f\left( {{W_2}, - 2\sqrt \gamma  } \right)}}} \right)} \right]\\
		{\rm{                   }} - \frac{1}{4}{\mathbb{E}_{W_2}}\left[ {\log \left( {1 + \frac{{f\left( {{W_2},\sqrt \gamma   + j\sqrt \gamma  } \right) + f\left( {{W_2},\sqrt \gamma   - j\sqrt \gamma  } \right)}}{{f\left( {{W_2},2\sqrt \gamma  } \right) + f\left( {{W_2},0} \right)}}} \right)} \right]\\
		{\rm{                   }} - \frac{1}{4}{\mathbb{E}_{W_2}}\left[ {\log \left( {1 + \frac{{f\left( {{W_2},\sqrt \gamma   - j\sqrt \gamma  } \right) + f\left( {{W_2}, - \sqrt \gamma   - j\sqrt \gamma  } \right)}}{{f\left( {{W_2},0} \right) + f\left( {{W_2}, - j2\sqrt \gamma  } \right)}}} \right)} \right]\\
		{\rm{                   }} - \frac{1}{4}{\mathbb{E}_{W_2}}\left[ {\log \left( {1 + \frac{{f\left( {{W_2},\sqrt \gamma   + j\sqrt \gamma  } \right) + f\left( {{W_2}, - \sqrt \gamma   + j\sqrt \gamma  } \right)}}{{f\left( {{W_2},j2\sqrt \gamma  } \right) + f\left( {{W_2},0} \right)}}} \right)} \right]
	\end{array}
\end{equation}
where ${W_2} \sim \mathcal{CN}(0,1)$ denotes a random variable for the two-dimensional AWGN, and the function $f(\cdot,\cdot)$ is given in \eqref{1-11}.

Since the high-level signal of the MLC and that of the OB-MLC are mathematically same in terms of modulations and demodulations, the difference of the mutual informations between the OB-MLC and that of the MLC can be calculate by 
\begin{equation}\label{2-7}
D_T(\bar{E}/\sigma^2) =  2I_o\left(\bar{E}/\sigma^2\right) - I_{LQ}\left(\bar{E}/\sigma^2\right)
\end{equation} 
where $D_T(\bar{E}/\sigma^2)$ denotes the difference of the two mutual informations at SNR $\bar{E}/\sigma^2$.

\begin{figure}
	\centering
	\includegraphics[width=0.5\textwidth]{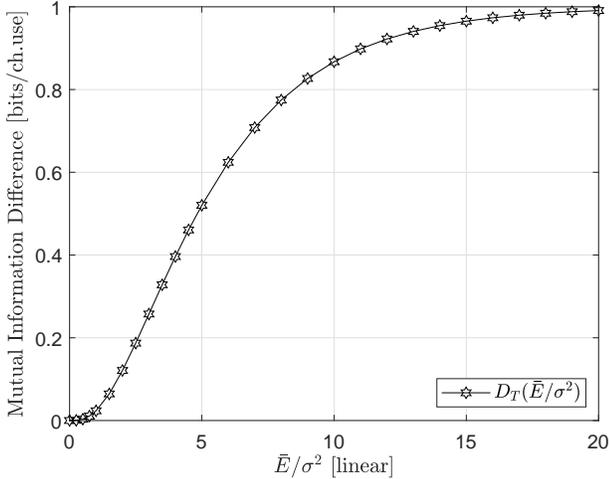}
	\caption{Difference between the mutual information of the two-dimensional OBAM and QPSK.}
	\label{fig_rate_2DL_diff}
\end{figure}

The numerical results of \eqref{2-7} are shown in Fig. \ref{fig_rate_2DL_diff}  for illustrating the gain.  It is clearly found that the gain is larger than zero in full range of the SNR and saturated at 1 bit/s/Hz which is the bandwidth efficiency gain of the OB-MLC.    

Finally, we note two points of the OB-MLC as follows.  First, the assumption of $K_c=K_o/2$ may not be exactly holds in practical applications. However, though $K_c$ can only be slightly different from $K_o$ relatively, its influence to bandwidth gain in \eqref{1-8} can be insignificant. (2) the weakness of OB-MLC is on the sensitivity to the error performance of the low-signals because CB demodulation requires the correctness of the OB on each bit value for finding each CB-symbol.  However, when the decode results are measure in terms of the error rate of the code word, the gain of OB-MLC holds theoretically.

\section{Conclusions}
This work leverages the concept OB method to organize a MLC signals of two levels for transmitting the OB and CB separately.  As a result, the bandwidth efficiency of the proposed method is increased by a factor of 1.5 in comparison with BPSK and QPSK in one- and two dimension.  Meanwhile, the mutual information is also increased in full rang of the SNR.  The numeral results conform this approach of the effectiveness.

\end{document}